\documentclass[aps,showpacs,pre,superscriptaddress]{revtex4}
\usepackage{epsfig}
\begin{document}
\title{Exact solutions of the Saturable Discrete Nonlinear Schr{\"o}dinger Equation}
\author{Avinash Khare}
\affiliation{Institute of Physics, Bhubaneswar, Orissa 751005, India}
\author{Kim~{\O}. Rasmussen}
\affiliation{Theoretical Division, Los Alamos National Laboratory,
Los Alamos, New Mexico, 87545, USA}
\author{Mogens R. Samuelsen}
\affiliation{Department of Physics, The Technical University of Denmark, DK-2800 Kgs. Lyngby, Denmark}
\author{Avadh Saxena}
\affiliation{Theoretical Division, Los Alamos National Laboratory,
Los Alamos, New Mexico, 87545, USA}
\date{\today}

\pacs{61.25.Hq, 64.60.Cn, 64.75.+g}

\begin{abstract}
Exact solutions to a nonlinear Schr{\"o}dinger lattice with a saturable 
nonlinearity are reported.
For finite lattices we find two different standing-wave-like 
solutions, and for an infinite lattice we find a
localized soliton-like solution. The existence requirements and stability 
of these solutions are discussed, and 
we find that our solutions are linearly stable in most cases. We also
show that the effective Peierls-Nabarro barrier potential is nonzero 
thereby indicating that this discrete model is quite likely nonintegrable. 
\end{abstract}
\maketitle

The discrete nonlinear Schr{\"o}dinger (DNLS) 
equation occurs ubiquitously \cite{KRB} throughout modern
science. Most notable is the role it plays in understanding 
the propagation of electromagnetic 
waves in glass fibers and other optical waveguides \cite{OP}. 
More recently it has been applied to describe 
Bose-Einstein condensates 
in optical lattices \cite{ST}. Here we are 
concerned with the DNLS equation with a saturable nonlinearity
\begin{equation}
i\dot{\psi}_n + (\psi_{n+1}+\psi_{n-1}-2\psi_n)+
\frac{\nu|\psi_n|^2}{1+\mu|\psi_n|^2}\psi_n=0,
\label{EQ:1}
\end{equation}
which is an established model for optical pulse propagation in various 
doped fibers \cite{fibers}. In Eq. (\ref{EQ:1}), $\psi_n$ is a complex valued
``wave function" at site $n$, while $\nu$ and $\mu$ are real 
parameters. This equation represents a Hamiltonian system with:  
\begin{equation}
{\cal H}= \sum_{n=1}^{N} \left [|\psi_n
-\psi_{n+1}|^2-\frac{\nu}{\mu}|\psi_n|^2
+\frac{\nu}{\mu^2} \ln \left (1+\mu|\psi_n|^2\right ) \right ],
\label{EQ:HAM}
\end{equation}
so that Eq. (\ref{EQ:1}) is given by 
$i \dot \psi_n =\frac{\partial{\cal H}}{\partial \psi_n^*}$. The dynamics of 
Eq. (\ref{EQ:1}) conserve, in addition to the Hamiltonian ${\cal H}$, the 
{\em power} ${\cal P}$
\begin{equation}
{\cal P}=\sum_{n=1}^{N} | \psi_n|^2.
\label{EQ:POWER}
\end{equation}
In the above equations $N$ is the number of lattice sites in the system. We note
that a transformation $\sqrt{\nu}\psi_n\rightarrow\psi_n$ will replace $\nu$ by
1 and $\mu$ by $\frac{\mu}{\nu}$ in the above equations. Note also
that Eq. (\ref{EQ:1}) is invariant under the transformation $\psi_n \rightarrow
\exp(i\delta) \psi_n$ where $\delta$ represents an abitrary phase.

For given system parameters $\nu$ and $\mu$ it can be shown, 
using recently derived \cite{Khare} local and
cyclic identities for Jacobi  elliptic functions \cite{stegun}, 
that Eq. (\ref{EQ:1}) 
has two (Case I and Case II) different temporally and spatially 
periodic solutions. Both solutions possess the temporal frequency
\begin{equation}
\omega=2\left(1-\frac{\nu}{2\mu}\right).
\label{EQ:Freq}
\end{equation}
Using standard notation \cite{stegun} for the Jacobi elliptic functions of modulus $m$ the solutions 
can be expressed as

Case I:

\begin{equation}
\psi_n^I=\frac{1}{\sqrt{\mu}}\frac{\mbox{sn}(\beta,m)}{\mbox{cn}(\beta,m)}
\mbox{dn}([n+c]\beta,m)
\exp \left (-i[\omega t+\delta] \right),
\label{EQ:dn}
\end{equation}
where the modulus $m$ must be chosen such that
\begin{equation}
\frac{2\mu}{\nu}=\frac{\mbox{cn}^2(\beta,m)}{\mbox{dn}(\beta,m)},
~~\beta=\frac{2K(m)}{N_p},
\label{EQ:CON_dn}
\end{equation}
and $c$ and $\delta$ are arbitrary constants. We only need to consider c between
0 and $\frac{1}{2}$ (half the lattice spacing).  Here $K(m)$ denotes 
the complete elliptic integral of first kind \cite{stegun}.  
While obtaining this solution, use has been made of the local identity
\begin{equation}
\mbox{dn}^2 (x,m)[\mbox{dn}(x+a,m)+\mbox{dn}(x-a,m)]=
-\frac{\mbox{cn}^2 (a,m)}{\mbox{sn}^2(a,m)}
[\mbox{dn}(x+a,m)+\mbox{dn}(x-a,m)]+2\frac{\mbox{dn}(a,m)}{\mbox{sn}^2(a,m)}
\mbox{dn} (x,m)\,,
\end{equation}
derived recently \cite{Khare}. In fact, given Eq. (\ref{EQ:1}) and
this local identity [and similar ones for $\mbox{sn}(x,m)$ and
$\mbox{cn}(x,m)$], it was straightforward to obtain the two solutions
presented here and the third solution follows simply by taking the 
limit $m \rightarrow 1$ of these two solutions as shown below.

\begin{figure}
\includegraphics[width=0.45\textwidth]{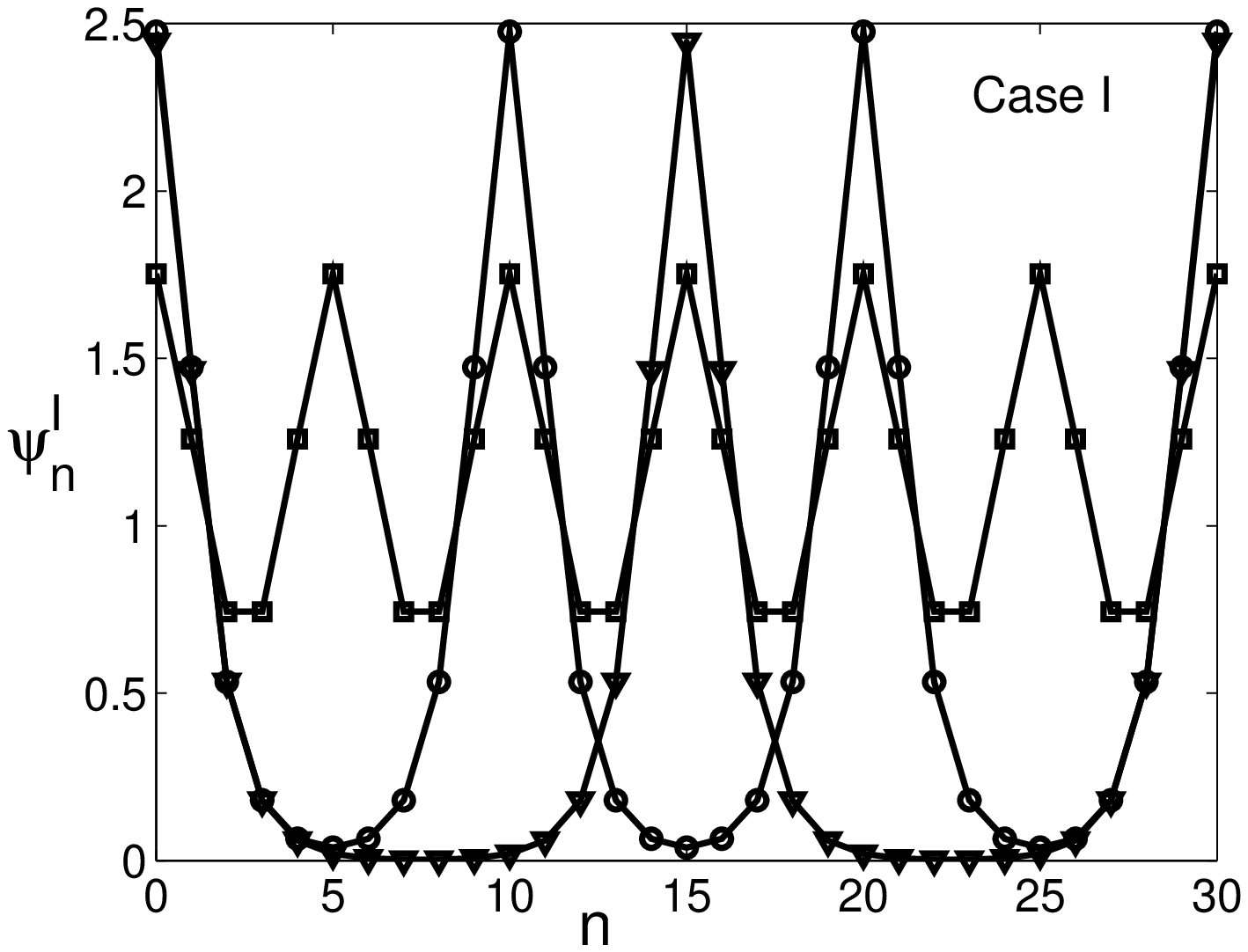}
\includegraphics[width=0.45\textwidth]{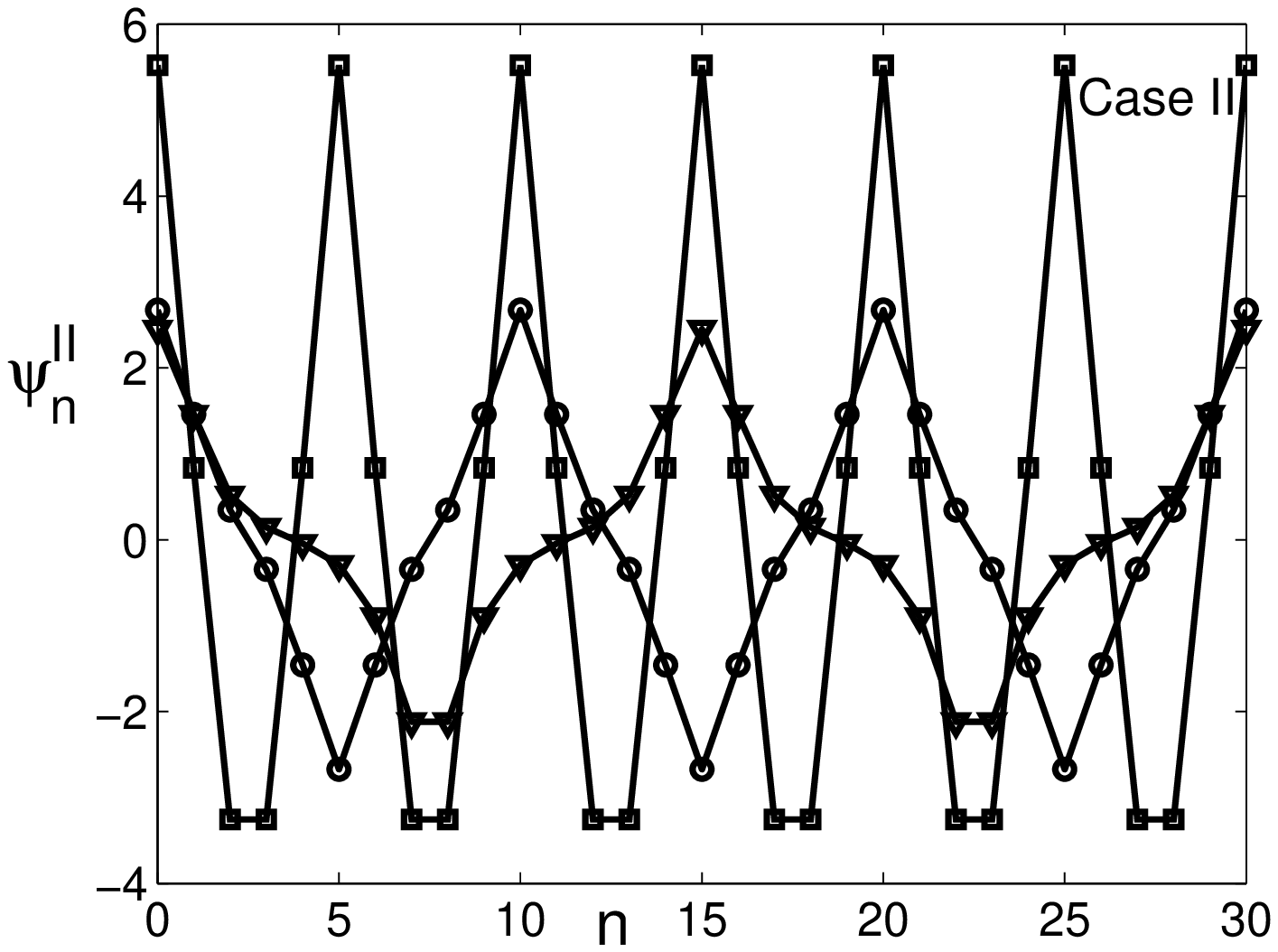}
 \caption{\label{fig:sol}Illustration of the exact solutions of two types. 
$\nu=1$, $\mu=0.3$, $\omega=-1.33$, and $c=t=\delta=0$. $N_p=5$ (squares), $N_p=10$ 
(circles), and $N_p=15$ (triangles). Lines are guides to the eye.
}
 \end{figure}

Case II:

\begin{equation}
\psi_n^{II}=\sqrt{\frac{m}{\mu}}\frac{\mbox{sn}(\beta,m)}{\mbox{dn}(\beta,m)}
\mbox{cn}([n+c]\beta,m)
\exp \left (-i[\omega t+\delta] \right),
\label{EQ:cn}
\end{equation}
where modulus $m$ now is determined such that
\begin{equation}
\frac{2\mu}{\nu}=\frac{\mbox{dn}^2(\beta,m)}{\mbox{cn}(\beta,m)},
~~\beta=\frac{4K(m)}{N_p}.
\label{EQ:CON_cn}
\end{equation}
While obtaining this solution, use has been made of the local identity
\cite{Khare} 
\begin{equation}
m \mbox{cn}^2 (x,m)[\mbox{cn}(x+a,m)+\mbox{cn}(x-a,m)]=
-\frac{\mbox{dn}^2 (a,m)}{\mbox{sn}^2(a,m)}
[\mbox{dn}(x+a,m)+\mbox{dn}(x-a,m)]+2\frac{\mbox{cn}(a,m)}{\mbox{sn}^2(a,m)}
\mbox{cn} (x,m)\,.
\end{equation}
Note that the two solutions, Eqs. (\ref{EQ:dn}) and (\ref{EQ:cn}), 
are translationally invariant. 

The two solutions $\psi_n^{I,II} $
are illustrated in Fig. \ref{fig:sol} for $t=\delta=c=0$.
In both cases the integer $N_p$ denotes the spatial period of the solutions. 
Both the solutions $\psi_n^I$ and $\psi_n^{II}$ reduce to the same 
localized solution in the limit $N_p\rightarrow\infty$ ($m\rightarrow 1$):

Case III:

\begin{equation}
\psi_n^{III}=\frac{1}{\sqrt{\mu}}\frac{\sinh(\beta)}{\cosh([n+c]\beta)}
e^{-i[\omega t+\delta]},
~~(N_p\rightarrow\infty),
\label{EQ:sech}
\end{equation}
where $\beta$ is now given by
\begin{equation}
\mbox{sech}\beta = \frac{2\mu}{\nu}.
\label{EQ:CON_sech}
\end{equation}
Again the frequency $\omega$ is given by Eq. (\ref{EQ:Freq}). This solution is noteworthy in that it 
is very similar in form to the celebrated exact soliton solutions of both the continuum cubic 
nonlinear Schr{\"o}dinger equation \cite{Drazin} and the (integrable) Ablowitz-Ladik lattice \cite{scott}

\begin{figure}[h]
\includegraphics[width=0.45\textwidth]{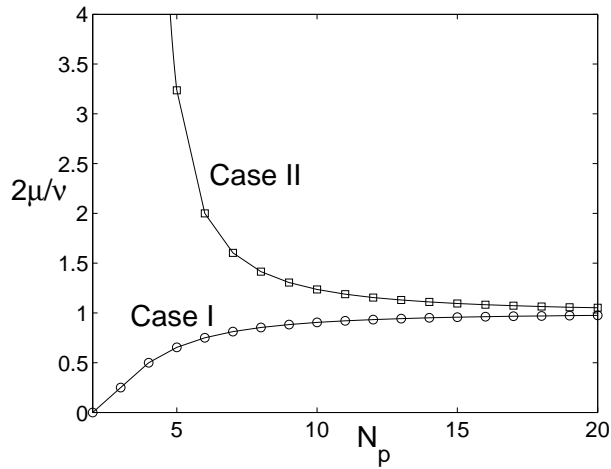}
 \caption{\label{fig:solFig2}Illustration of parameter values $\mu$, $\nu$,
and $N_p$ for which the exact solutions are allowed. Case I: 
$2\mu/\nu$ between 0 and $\cos^2\frac{\pi}{N_p}$ and $N_p\geq 3$. Case II:
$2\mu/\nu$ between 0 and $1/\cos^2\frac{2\pi}{N_p}$ and $N_p\geq 3$ except for
$N_p=4$.
}
\end{figure}

There are, as expressed by Eqs. (\ref{EQ:CON_dn}), (\ref{EQ:CON_cn}), and 
(\ref{EQ:CON_sech}), stringent conditions on the parameters $\mu$ and $\nu$ 
for which these exact solutions exist. In the cases I and II these limitations are illustrated 
in Fig. \ref{fig:solFig2}, which shows that the solution $\psi_n^{I}$ only exists for parameter values 
below the lower curve (circles). Similarly, the  solution $\psi_n^{II}$ for periods $N_p>4$ only exists 
below the upper curve (squares). As can be easily seen from Eq. (\ref{EQ:CON_cn}) the  $\psi_n^{II}$ 
solution does not exist for $N_p=4$.  However, it does exist for $N_p=3$, but only for parameter ratios $\mu/\nu <0$.
As a result of the periodic boundary conditions both solutions become meaningless for $N_p<3$.
The solution  $\psi_n^{III}$ exists for all parameter values $\nu \geq 2 \mu > 0
$.

For the $\psi_n^{III}$ solution, expressions for both the power Eq. 
(\ref{EQ:POWER}) and the Hamiltonian Eq. (\ref{EQ:HAM}) can be obtained 
by using exact (Poisson) summation rules \cite{Avadh}
\begin{equation}
{\cal P}^{III}
=\frac{2}{\mu}
\frac{\mbox{sinh}^2(\beta)}{\beta^2}
\big[\beta-2K(m)E(m)+2K^2(m)\mbox{dn}^2(2K(m)c,m) \big], 
\label{EQ:POWER_III}
\end{equation}
\begin{equation}
{\cal H}^{III}=-\frac{4}{\mu}\mbox{sinh}(\beta)
+\left(1-\frac{\nu}{2\mu}\right)
\frac{4}{\mu}\frac{\mbox{sinh}^2(\beta)}{\beta^2}
\big[\beta-2K(m)E(m)+2K^2(m)\mbox{dn}^2(2K(m)c,m) \big]  
+\frac{\nu}{\mu^2}2\beta.
\label{EQ:HAM_III}
\end{equation}
Here the modulus $m$ must be determined such that 
\begin{equation}
\beta=\pi\frac{K(m)}{K(m_1)},~~\mbox{sech}\beta=\frac{2\mu}{\nu}, 
\end{equation}
where $m_1=1-m$ is the complementary modulus and $E(m)$ denotes the complete 
elliptic integral of the second kind.  For the cases I and II analogous  
expressions can be obtained and they are given in the Appendix.

In a discrete lattice there is an energy cost associated with moving a 
localized mode (such as a soliton or a breather) by a half lattice constant.  
This is called the Peierls-Nabarro (PN) barrier \cite{PN,peyrard}.  Having 
obtained the expression for ${\cal H}^{III}$ analytically in a closed form, 
we can now calculate the energy difference between the solutions when $c=0$ 
and $c=1/2$, i.e. when the peak of the solution is centered on a lattice 
site and when it is centered half-way between two adjacent sites, 
respectively. We find that
\begin{equation}
\Delta E \equiv {\cal H}^{III} (c=0) - {\cal H}^{III} (c=1/2) = - \frac{16m} 
{\mu \beta^2} \sinh^2 (\beta) \sinh^2 (\beta /2)K^2(m) < 0\,,
\end{equation}
that is, the energy is lowest when the peak of the solution is centered
at the sites. Thus, there is a finite energy barrier (i.e. the height of 
the effective PN barrier potential) between these two stationary states 
due to discreteness. If the folklore of nonzero PN barrier being 
indicative of non-integrability of the discrete nonlinear system is correct, 
this suggests that quite likely our discrete model is non-integrable 
unlike the Ablowitz-Ladik model \cite{scott}.  
 
In order to study the linear stability of the exact solutions $\psi_n^j$ 
($j$ is I, II, or III) we introduce the following expansion
\begin{equation}
\psi_n(t)=\psi_n^j+\delta \psi_n(t) e^{-i\omega t},
\label{EQ:STAB_1}
\end{equation}
applied in a frame rotating with frequency $\omega$ of the solution. 
Substituting into Eq. (\ref{EQ:1}) and retaining only terms linear in the 
perturbation we get
\begin{equation}
i\delta\dot{\psi}_n+\big(\delta \psi_{n+1}+\delta \psi_{n-1}-2\delta \psi_n\big)+
\left(\omega +\frac{\nu|\psi_n^j|^2(2+\mu|\psi_n^j|^2)}{(1+\mu|\psi_n^j|^2)^2}\right)
\delta \psi_n+
\frac{\nu|\psi_n^j|^2}{(1+\mu|\psi_n^j|^2)^2}\delta \psi_n^*
=0.
\label{EQ:STAB_LIN}
\end{equation}
Continuing by splitting the perturbation $\delta \psi_n $ into real parts $\delta u_n$
and imaginary parts $\delta v_n$ ($\delta \psi_n =\delta u_n+i\delta v_n$)
and introducing the two real vectors
\begin{eqnarray}
\delta\mbox{\boldmath $U$}=\{\delta u_n\}&~~\mbox{and}~~&
\delta\mbox{\boldmath $V$}=\{\delta v_n\}
\end{eqnarray}
and the two real matrices $\mbox{\boldmath $A$}=\{A_{nm}\}$ and
$\mbox{\boldmath $B$}=\{B_{nm}\}$ by defining
\begin{eqnarray}
A_{nm}&=&
\delta_{n,m+1}+\delta_{n,m-1}+
\left (\omega-2+\frac{\nu|\psi_n^j|^2(3+\mu|\psi_n^j|^2)}{(1+\mu|\psi_n^j|^2)^2}\right )
\delta_{nm},\\
B_{nm}&=&
\delta_{n,m+1}+\delta_{n,m-1}+
\left (\omega-2+\frac{\nu|\psi_n^j|^2}{(1+\mu|\psi_n^j|^2)}\right )
\delta_{nm},
\end{eqnarray}
where $m\pm 1$ in the Kronecker $\delta$ means: $m\pm 1~mod~N$.  Then Eq. 
(\ref{EQ:STAB_LIN}) can be written compactly as 
\begin{eqnarray}
-\delta\mbox{\boldmath $\dot{V}$}+
\mbox{\boldmath $A$}\delta\mbox{\boldmath $U$}=\mbox{\boldmath $0$},&\mbox{and}&
\delta\mbox{\boldmath $\dot{U}$}+
\mbox{\boldmath $B$}\delta\mbox{\boldmath $V$}=\mbox{\boldmath $0$}, 
\end{eqnarray}
where an overdot denotes time derivative.  Combining these first order differential equations we get:
\begin{eqnarray}
\delta\mbox{\boldmath $\ddot{V}$}+\mbox{\boldmath $A$}\mbox{\boldmath $B$}
\delta\mbox{\boldmath $V$}=\mbox{\boldmath $0$},&\mbox{and}&
\delta\mbox{\boldmath $\ddot{U}$}+\mbox{\boldmath $B$}\mbox{\boldmath $A$}
\delta\mbox{\boldmath $U$}=\mbox{\boldmath $0$}.
\end{eqnarray}
The two matrices $\mbox{\boldmath $A$}$ and $\mbox{\boldmath $B$}$ are
symmetric and have real elements. However, since they do not commute
$\mbox{\boldmath $A$}\mbox{\boldmath $B$}$ and
$\mbox{\boldmath $B$}\mbox{\boldmath $A$}=
(\mbox{\boldmath $A$}\mbox{\boldmath $B$})^{T}$
are not symmetric.
$\mbox{\boldmath $A$}\mbox{\boldmath $B$}$ and
$\mbox{\boldmath $B$}\mbox{\boldmath $A$}$ have the same eigenvalues, but different
eigenvectors. The eigenvectors for each of the two
matrices need not be orthogonal.
\begin{figure}
\includegraphics[width=0.45\textwidth]{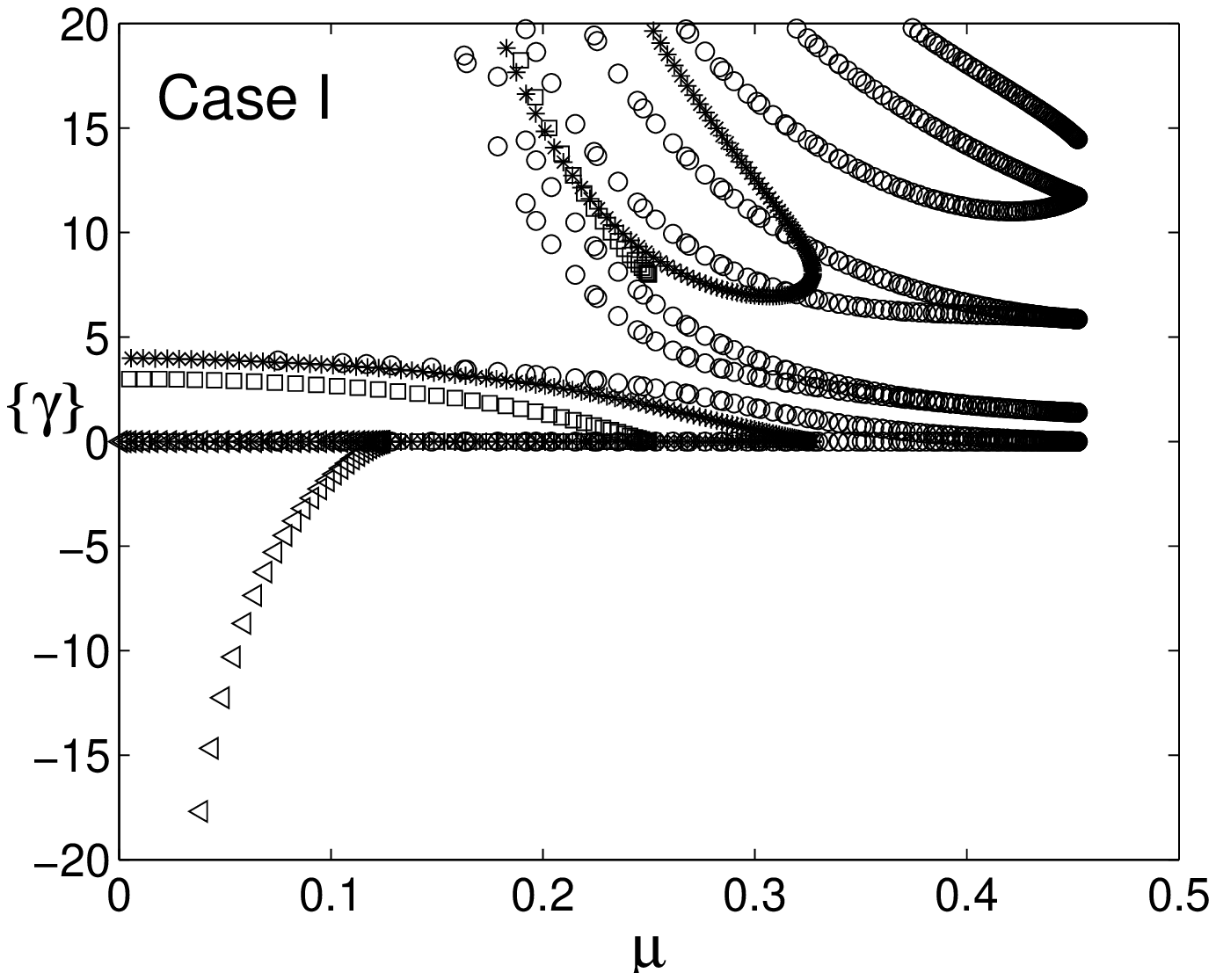}
\includegraphics[width=0.45\textwidth]{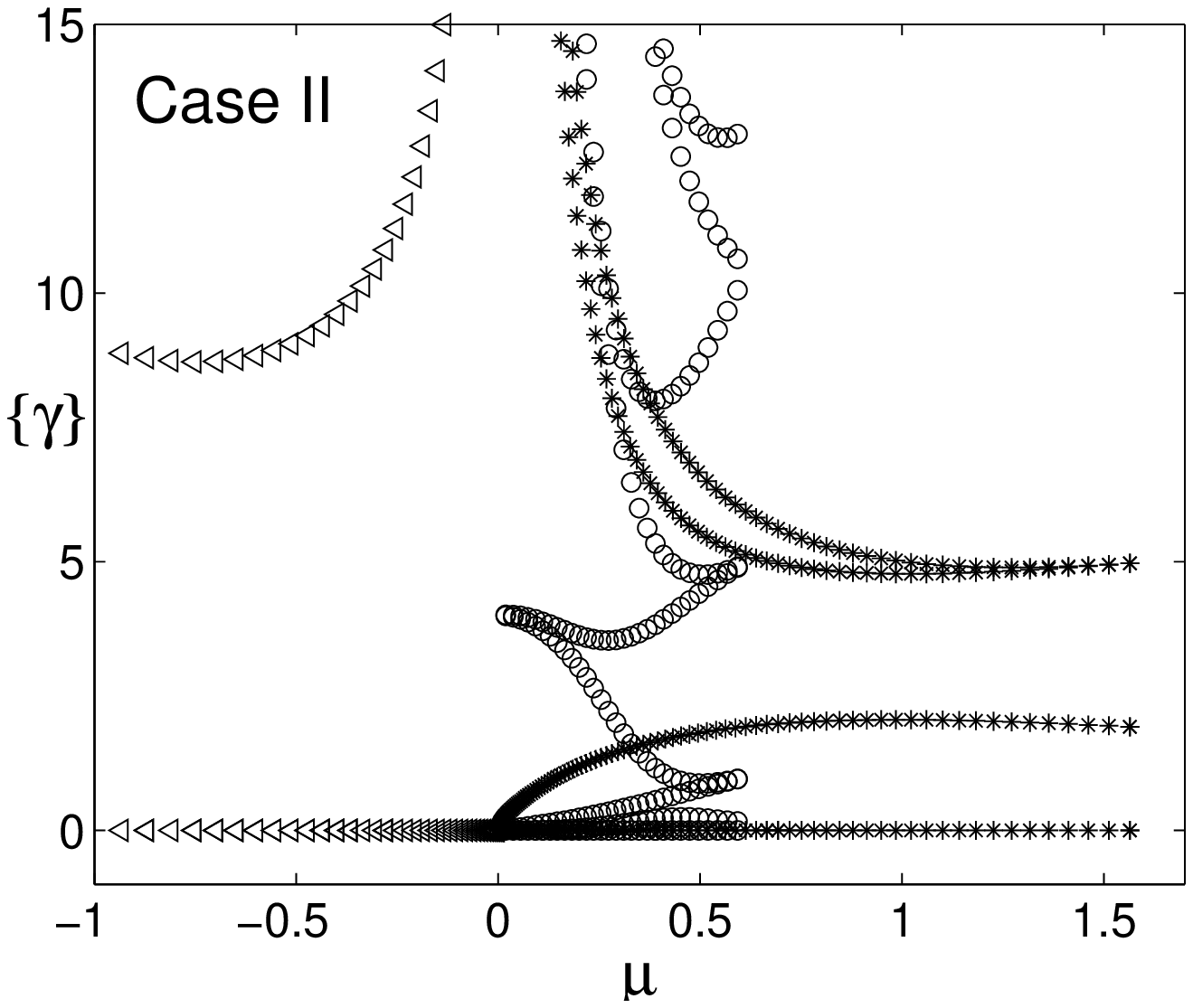}
 \caption{\label{FIG:SPEC}Illustration of the stability of the exact 
solutions.  Shown is the eigenvalue spectrum $\{\gamma\}$ for the matrix 
product $\mbox{\boldmath $A$} \mbox{\boldmath $B$}$, $\nu=1$. Case I 
(left panel) and $N_p=3$ (triangles), $N_p=4$ (squares), $N_p=5$ (stars), 
and $N_p=10$ (circles). Case II (right panel) $N_p=3$ (triangles), 
$N_p=5$ (stars), and $N_p=10$ (circles). 
}
 \end{figure}

\noindent

The eigenvalue spectrum $\{\gamma \}$ of the matrices $\mbox{\boldmath $A$}\mbox{\boldmath $B$}$ and
$\mbox{\boldmath $B$}\mbox{\boldmath $A$}$ determines the stability of the
exact solutions. If it contains negative eigenvalues the solution is
unstable. The eigenvalue spectrum  always contains two eigenvalues which are zero. These eigenvalues
correspond to the translational invariance ($c$) 
and to the invariance of the solution $\psi_n^j$ to a constant phase factor
$e^{-i\delta}$ (i.e. translation in time), respectively. In Fig. \ref{FIG:SPEC} we show the eigenvalue spectrum $\{\gamma\}$ for the cases I and II for several
periodicities $N_p$. It is important to note that in this figure we have $N=N_p$.
It turns out that the spectrum $\{\gamma\}$ is independent of $c$.
The figure demonstrates that 
for $N=N_p$, only the $\psi_n^{I}$ solution becomes unstable and this occurs only for $N_p=3$. For all 
other values of $N_p$ both solutions are linearly stable. This also indicates that the localized 
solution $\psi_n^{III}$ is linearly stable; and we have checked that this indeed is the case in
the entire existence interval.

The solutions $\psi_n^{I}$, and $\psi_n^{II}$ exist for all lattices $N=JN_p$ 
where $J$ is a positive integer. However, we find $\psi_n^{I}$ to be stable 
only for $J=1$, while $\psi_n^{II}$ is stable for all $J$.

Finally, it is worth pointing out that 
Eq. (\ref{EQ:1}) also has an exact constant amplitude solution
\begin{equation}
\psi_n (t) = \psi_0 \exp [-i(\omega t-qn+\delta)]\,,
\end{equation}
where $\delta$ is a constant and $\omega$ satisfies the {\it
nonlinear} dispersion relation
\begin{equation}
\omega = 4\sin^2 (q /2) - \frac{\nu \mid \psi_0 \mid^2}{1+\mu \mid
\psi_0 \mid^2}\,,
\end{equation}
where the wavenumber $q=2\pi p/N_p$ in order to comply with the periodic boundary condition, and $p$ is 
an intger. 

In conclusion, we have presented two spatially periodic and one 
spatially localized exact solutions of the DNLS equation with a 
saturable nonlinearity.  We found these solutions to be linearly 
stable in most cases.  We also calculated the Peierls-Nabarro 
barrier for the localized solution.  These results are relevant 
for wave propagation in optical waveguides and doped fibers 
\cite{OP,fibers}, Bose-Einstein condensates \cite{ST} as well as 
for many other nonlinear physical applications.  Note that a related 
continuum version of Eq. (1), which arises in the context of the 
Fokker-Planck equation for a single mode laser, has been considered 
in Ref. \cite{SH}. It would be important to search for ways of 
modifying the nonlinearity so that the PN barrier becomes zero--a 
possible route to an integrable model.\\

This work was supported in part by the U.S. Department of Energy. 
 
\appendix
\section{}
In this appendix we give explicit expressions for $H$ and $P$ for the
two spatially periodic solutions. While the importance of the energy
expression is obvious, we would like to emphasize that the expressions
for $P$ could be used as a numerical diagnostic, for instance in keeping 
track of a conserved quantity in a simulation involving these solutions.

Inserting the solution given by Eq. (\ref{EQ:dn}) into Eq. (\ref{EQ:HAM}) we get
for the energy
\begin{eqnarray}
{\cal H}^I=\frac{2}{\mu}\frac{\mbox{sn}^2(\beta,m)}{\mbox{cn}^2(\beta,m)}
\bigg(
-N_p\big(\mbox{dn}(\beta,m)-\mbox{cs}(\beta,m)Z(\beta,m)\big)
+\left(1-\frac{\nu}{2\mu}\right)
\sum_{n=1}^{N_p} \mbox{dn}^2 ([n+c]\beta,m)
\bigg )\nonumber\\
+\frac{\nu}{\mu^2}
\sum_{n=1}^{N_p} \mbox{ln}\left(1+\frac{\mbox{sn}^2(\beta,m)}{\mbox{cn}^2(\beta,m)}
\mbox{dn}^2([n+c]\beta,m)\right),
\label{EQ:HAM_I}
\end{eqnarray}
where $Z(\beta,m)$ is the Jacobi zeta function and cs$(\beta,m)$ =
cn$(\beta,m)$/sn$(\beta,m)$. Also, use has been made of the identity \cite{Khare}
dn$(y,m)$dn$(y+a,m)$=dn$(a,m)-$cs$(a,m)$Z$(a,m)$+cs$(a,m)$[Z$(y+a,m)$-Z$(y,m)]$ 
and the fact that $\sum_{n=1}^{N_p}[Z(\beta(n+1+c),m)-Z(\beta(n+c),m)]=0$.
From Eq. (\ref{EQ:POWER}) we get for the power
\begin{equation}
{\cal P}^I =
\frac{1}{\mu}
\frac{\mbox{sn}^2(\beta,m)}{\mbox{cn}^2(\beta,m)}
\sum_{n=1}^{N_p} \mbox{dn}^2([n+c]\beta,m).
\label{EQ:POWER_I}
\end{equation}
Similarly, inserting the solution given by Eq. (\ref{EQ:cn}) into Eq. 
(\ref{EQ:HAM}) we get for the energy

\begin{eqnarray}
{\cal H}^{II}
=\frac{2}{\mu}\frac{\mbox{sn}^2(\beta,m)}{\mbox{dn}^2(\beta,m)}
\bigg(
-N_p\big(m\,\mbox{cn}(\beta,m)-\mbox{ds}(\beta,m)Z(\beta,m)\big)
+\left(1-\frac{\nu}{2\mu}\right)
\sum_{n=1}^{N_p} \mbox{cn}^2 ([n+c]\beta,m)
\bigg )\nonumber\\
+\frac{\nu}{\mu^2}
\sum_{n=1}^{N_p} \mbox{ln}\left(1+\frac{\mbox{sn}^2(\beta,m)}{\mbox{dn}^2(\beta,m)}
\mbox{cn}^2([n+c]\beta,m)\right)\nonumber\\
~~~~~\nonumber\\
=\frac{2}{\mu}\frac{\mbox{sn}^2(\beta,m)}{\mbox{dn}^2(\beta,m)}
\bigg( -N_p\big[m\,\mbox{cn}(\beta,m)-\mbox{ds}(\beta,m)Z(\beta,m)\big] 
+\left(1-\frac{\nu}{2\mu}\right)\big[-(1-m)N_p+\sum_{n=1}^{N_p}
\mbox{dn}^2 ([n+c]\beta,m)\big]\bigg)\nonumber\\
+\frac{\nu}{\mu^2}\bigg (
N_p\mbox{ln}\left(\frac{\mbox{cn}^2(\beta,m)}{\mbox{dn}^2(\beta,m)}\right)
+\sum_{n=1}^{N_p} \mbox{ln}\left[1+\frac{\mbox{sn}^2(\beta,m)}{\mbox{cn}^2(\beta,m)}
\mbox{dn}^2([n+c]\beta,m)\right]\bigg ),
\label{EQ:HAM_II}
\end{eqnarray}
where again $Z(\beta,m)$ is the Jacobi zeta function and ds$(\beta,m)$ =
dn$(\beta,m)$/sn$(\beta,m)$. Also, use has been made of the identity \cite{Khare}
$m\,$cn$(y,m)$cn$(y+a,m)$=$m\,$cn$(a,m)-$ds$(a,m)$Z$(a,m)$+ds$(a,m)$
[Z$(y+a,m)$-Z$(y,m)$].
From Eq. (\ref{EQ:POWER}) we get for the power
\begin{eqnarray}
{\cal P}^{II}
=\frac{1}{\mu}
\frac{\mbox{sn}^2(\beta,m)}{\mbox{dn}^2(\beta,m)}
\sum_{n=1}^{N_p} \mbox{cn}^2([n+c]\beta,m)  
=\frac{1}{\mu}
\frac{\mbox{sn}^2(\beta,m)}{\mbox{dn}^2(\beta,m)}
\left(-N_p(1-m)+\sum_{n=1}^{N_p} \mbox{dn}^2([n+c]\beta,m)\right).
\label{EQ:POWER_II}
\end{eqnarray}
In order to get the sums over the same expressions for ${\cal H}^{II}$ and 
${\cal P}^{II}$ as for ${\cal H}^{I}$ and ${\cal P}^{I}$ we have used the 
basic relations $\mbox{cn}^2 (x,m)+\mbox{sn}^2 (x,m)=1$ and $\mbox{dn}^2 
(x,m) +m\mbox{sn}^2 (x,m) =1$.  In the continuum limit (small $\beta$, 
large $N_p$) the sums may be replaced by integrals. First
\begin{equation}
\sum_{n=1}^{N_p} \mbox{dn}^2 ([n+c]\beta,m)\simeq\frac{QE(m)}{\beta}=
\frac{QK(m)}{\beta}\frac{E(m)}{K(m)}=N_p\frac{E(m)}{K(m)},
\label{EQ:SUM_1}
\end{equation}
where $Q= 2$ in Case I and $Q=4$ in Case II. The other sum
\begin{equation}
\sum_{n=1}^{N_p} \mbox{ln}\left(1+\frac{\mbox{sn}^2(\beta,m)}{\mbox{cn}^2(\beta,m)}
\mbox{dn}^2([n+c]\beta,m)\right)\simeq N_p\mbox{ln}
\bigg(\frac{\pi\Theta^2(\beta,m)}{2\sqrt{1-m}K(m)\mbox{cn}^2(\beta,m)}\bigg),
\label{EQ:SUM_@}
\end{equation}
where $\Theta(\beta,m)$ is the Jacobi theta function.  For $m\rightarrow 1$, 
Eqs. (A1) and (A3) can be used to determine the asymptotic interaction 
between two nonlinear solutions given by Eq. (11). \\

\end{document}